# Highly accelerated MR parametric mapping by undersampling the k-space and reducing the contrast number simultaneously with deep learning

Yanjie Zhu[1†], Haoxiang Li[1,2,3†], Yuanyuan Liu[1,2,3], Muzi Guo[1,2,3], Guanxun Cheng[5], Gang Yang[4], Haifeng Wang[1], Dong Liang[1*]

[1]Paul C. Lauterbur Research Center for Biomedical Imaging, Shenzhen Institutes of Advanced Technology, Shenzhen 518055, China

[2]University of Chinese Academy of Sciences, Beijing, China

[3]Shenzhen College of Advanced Technology, University of Chinese Academy of Sciences, Shenzhen, Guangdong, China

[4]Jinling Hospital, Medical School of Nanjing University, Nanjing, Jiangsu, China

[5]Peking University Shenzhen Hospital, Peking University, Shenzhen, Guangdong, China

# ABSTRACT

**Purpose:** To propose a novel deep learning-based method called RG-Net (reconstruction and generation network) for highly accelerated MR parametric mapping by undersampling k-space and reducing the acquired contrast number simultaneously.

**Methods:** The proposed framework consists of a reconstruction module and a generative module. The reconstruction module reconstructs MR images from the acquired few undersampled k-space data with the help of a data prior. The generative module then synthesizes the remaining multi-contrast images from the reconstructed images, where the exponential model is implicitly incorporated into the image generation through the supervision of fully sampled labels. The RG-Net was evaluated on the $T_{1\rho}$ mapping data of knee and brain at different acceleration rates. Regional $T_{1\rho}$ analysis for cartilage and the brain was performed to access the performance of RG-Net.

**Results:** RG-Net yields a high-quality $T_{1\rho}$ map at a high acceleration rate of 17. Compared with the competing methods that only undersample k-space, our framework achieves better performance in $T_{1\rho}$ value analysis. Our method also improves quality of $T_{1\rho}$ maps on patient with glioma.

**Conclusion:** The proposed RG-Net that adopted a new strategy by undersampling k-space and reducing the contrast number simultaneously for fast MR parametric mapping, can achieve a high acceleration rate while maintaining good reconstruction quality. The generative module of our framework can also be used as an insert module in other fast MR parametric mapping methods.

## 1 | INTRODUCTION

Quantitative magnetic resonance (MR) parametric mapping, such as $T_1$, $T_2$, and $T_{1\rho}$ relaxations, is a powerful tool for assessing tissue properties in the diagnosis and prognosis of diseases [1-6]. However, parametric mapping requires the acquisition of multiple images with different contrast weightings, leading to a long scan time that greatly hinders its wide clinical use [7,8]. The total scan time is equal to the scan time for a single-contrast image multiplied by the contrast number in the parametric direction. Therefore, there could be two strategies to reduce the scan time of parametric mapping: undersampling the k-space data of each image and decreasing the contrast number.

The prevailing methods for fast MR parametric mapping only use the first strategy. They undersample the k-space data first and then reconstruct the weighted images or the parametric maps with compressed sensing (CS)-based algorithms [9-26]. These methods can be classified into three categories; for details, please refer to [16]. Recently, several deep learning-based methods for fast MR parametric mapping have been developed [27-31]. These methods use end-to-end convolutional neural networks to reconstruct weighted images or direct estimate parametric maps from undersampled k-space data. Although these methods explored the correlation between multi-contrast images well, the achievable acceleration factor is still limited because only one strategy is used [32]. Therefore, it is highly desirable to combine two strategies for achieving a higher acceleration rate. However, little work has explored the possibility of reducing the contrast number in previous research on fast MR parametric mapping. Li et al[33] developed a deep learning method to directly estimate the parametric maps from undersampled k-space of reduced echoes. Theoretically, as long as the acquired contrast number is equal to the unknown number in the physical model of MR relaxation, the relaxation can be well estimated by pixel-wise curve fitting from the images. Practically, the fitting process needs to be overdetermined to overcome the uncertainty introduced by noise contamination, signal destabilization, hardware imperfections, etc. [34,35]. The needed contrast number is hence much larger than the unknown number. For example, $T_{1\rho}$ mapping with single exponential decay model has

2 unknowns but usually needs 5 to 8 $T_{1\rho}$-weighted images [19,36,37].

Therefore, there is a gap between reducing the contrast number for a short scan time and keeping it for a well-estimated parametric map. One possible solution to deal with the combined strategy is to synthesizes the remaining multi-contrast images from few acquired k-space data, which can reduce the scan time while maintaining the contrast number. Such generating methods can hardly be established with the acquired data itself and require big historical data for exploration. Deep learning methods have a strong potential for multi contrast images generation [38,39].

In this work, we propose a deep learning-based method with a reconstruction network and a generative network to highly accelerate MR parametric mapping by two strategies. The reconstruction network is to recover information loss in k-space undersampling and the generative network is to recover information loss in the dimension of contrast. More specifically, only two undersampled k-spaces with different contrasts are acquired, and the corresponding contrast-weighted images are then reconstructed via a reconstruction network, followed by generative network mapping of other desired contrast-weighted images from these two images. The generated images are constrained by the structural similarity between multi-contrast images and the relaxation model for information recovery. We took $T_{1\rho}$ mapping as an example to evaluate the performance of the proposed method. Extensive experiments were performed on three-dimensional (3D) $T_{1\rho}$ mapping data of the knee and brain. The results show that the proposed method can achieve a very high acceleration factor up to 17 for the knee and brain in $T_{1\rho}$ mapping with a lower nRMSE than the strategy that only undersamples the k-space.

## 2 | METHODS

Our framework for fast MR parametric mapping consists of two parts: a reconstruction module and a generative module (see Figure 1(a)). Both modules are designed based on a deep learning network and are described in detail next.

### 2.1 Reconstruction module

In fast MR imaging, CS is a popular method for accelerating data acquisition with the help of data prior knowledge, such as sparsity and low rank [11,12,32]. The general CS-MRI reconstruction formula can be described as:

$$\hat{m} = \underset{m}{argmin} \frac{1}{2} \|Am - y\|_2^2 + \lambda \|\psi m\|_p \tag{1}$$

where m is the MR image to be reconstructed, y is the corresponding multi-channels undersampled k-space data, and $A = \mathbb{p}FC$ is the measurement matrix. $\mathbb{p}$ is the undersampling mask, F is the Fourier transform and C is the coil sensitivity maps. $\|\psi m\|_p$ denotes the regularization term, which can enforce the prior information that improves image quality for undersampled data. In CS, the sparsity prior is usually used as the regularization term, where ψ is the sparsifying transform.

We applied ADMM-Net III[40,41] as the reconstruction module because it's a model-based unrolling method and shows high-quality reconstruction results from few undersampled data in our previous study. ADMM-Net III is a generalized version of ADMM-Net[39,42]. The solution iterations can be written as:

$$\begin{cases} D^{(n)}: d^{(n)} = \Gamma(Am^{(n-1)}, y) \\ M^{(n)}: m^{(n)} = \Pi(m^{(n-1)}, z^{(n-1)} - \beta^{(n-1)}, A^T d^{(n)}) \\ Z^{(n)}: z^{(n)} = \Lambda(m^{(n)}, \beta^{(n-1)}) \\ P^{(n)}: \beta^{(n)} = \beta^{(n-1)} + \tilde{\eta}(m^{(n)} - z^{(n)}) \end{cases} \tag{2}$$

The operators Γ, Π, Λ and the parameter η̃ are all learned by the network. In contrast, only the parameters and the priori regularization are learned in the original ADMM-Net[39]. In this work, we set the maximum iteration number as N = 10 empirically. Each iteration block consists of three operators (Γ, Π, Λ) and a learnable update parameter η̃, and the operators Γ, Π, Λ are replaced by learnable convolution layers and activation function layers. As shown in Figure 1(b), the reconstruction module consists of 10 iteration blocks, and the data flow of each iteration block is shown on the right side.

We take the fully sampled images as the ground truth $m^{gt}$ and the undersampled k-space data y as the input. Then, we construct a training dataset containing K pairs of undersampled k-space data and ground-truth images. We choose the normalized root mean square error (nRMSE) as the loss function of the reconstruction module, and the loss between the network output and ground truth is defined as:

$$Loss_1 = \frac{1}{K}\sum_{i=1}^{K}\frac{\sqrt{\left\|R(y_i|\theta)-m_i^{gt}\right\|_2^2}}{\sqrt{\left\|m_i^{gt}\right\|_2^2}} \qquad (3)$$

where $R(y_i|\theta)$ is the network output conditioned on network parameter θ and $i$th undersampled k-space data $y_i$. The network parameter θ is optimized by minimizing the above loss to reconstruct images from the undersampled k-space data.

**2.2 Generative network**

We propose a generative module with a densely connected convolutional neural network structure to generate unacquired contrast-weighted images from the few reconstructed images. The structure of the generative module is shown in Figure 1(c). There is a total of five Convblocks in the module. Each Convblock consists of 3 convolution layers with kernel sizes of 3*3 and 2 ReLU layers. The last four Convblocks are connected behind the channel-wise concatenation. Densely connected shortcut structures are helpful for probing robust and spatially invariant image features, which have shown better performance for image translation tasks in recent studies [43-45]. The loss between the results of this end-to-end generative module and fully sampled label can be expressed as:

$$Loss_2 = \frac{1}{JK}\sum_{i=1}^{K}\sum_{j=1}^{J}\left\|G(r_i|\theta)_j - s_{ij}\right\|_2 \qquad (4)$$

Here, $G(r_i|\theta)$ denotes the generated synthetic multi-contrast images with network parameter θ and input data $r_i$. $r_i$ is a 3D matrix of stacked two-dimensional (2D) images with a reduced contrast number. $s_i$ is a 3D matrix of stacked 2D images with the full contrast number minus the reduced contrast number (a total of $J$ channels). The corresponding fully sampled contrast images are used for training. $K$ is the number of image pairs belonging to the training dataset for the generative module. Similar to most deep learning-based image translation works, the $l_2$ norm is selected as a loss function to ensure the uniformity between the output images and ground truth [46-49].

**2.3 Data acquisition**

All experiments were performed on a UIH 3.0T scanner (uMR 790, United Imaging, Shanghai, China) and were approved by the local Institutional Review Board (IRB). Informed consent was obtained from each volunteer prior to the scan. Two fully sampled datasets were acquired with a 3D $T_{1\rho}$ mapping sequence with a spin-lock frequency of 500 Hz [50]. The first was a series of 12 channels knee $T_{1\rho}$-weighted images from 8 volunteers (4 males and 4 females, age: 25 ± 4 years). The imaging parameters were as follows: TE/TR = 8.96/2000 ms, field of view (FOV) = 256×146×124, matrix size = 256×146×124, echo train length = 60, spin-lock times (TSLs) = 5, 10, 20, 40, and 60 ms. The second was a series of 32 channels brain $T_{1\rho}$-weighted images from 8 volunteers (4 males and 4 females, age: 24 ± 3 years) and a 11 channels brain $T_{1\rho}$-weighted images from one patient with glioma. Due to the limitations of the patient's physical conditions and the different coils used in data acquisition, the data was acquired with acceleration factor of k-space $R_K$ = 3 and reconstructed on a UIH 3.0T scanner rather than using trained RG-Net. The imaging parameters were as follows: TE/TR = 7.84/2400 ms, FOV = 240×216×86, matrix size = 240×216×86, echo train length = 60, TSLs = 1, 15, 25, 45, and 65 ms. The total scan time of each volunteer is 51 mins 55 seconds (10 mins 23 seconds*5) for the brain and 48 mins 45 seconds (9 min 45 seconds *5) for the knee. The $T_{1\rho}$-weighted images acquired at each TSL are referred to as TSL1, TSL2, TSL3, TSL4, and TSL5 for brevity.

The reference images were obtained by applying an inverse fast Fourier transform (IFFT) to the fully sampled k-space data followed by coil sensitivity weighted combination. An adaptive method was used to calculate the coil sensitivity [51]. The fully sampled k-space data at each TSL were retrospectively undersampled by a Poisson variable density sampling mask [52] with acceleration factors in k-spaces ($R_k$) of 4.6 and 6.8. The sampling masks for each TSL were different according to the CS theory [53].

The $T_{1\rho}$ map was estimated pixel-by-pixel with the nonlinear least squares fitting method of the $T_{1\rho}$-weighted images according to the mono-exponential decay model shown as follows.

$$S(TSL_i) = S_0 e^{-\frac{TSL_i}{T_{1\rho}}}, i = 1, 2, 3, 4, \text{ and } 5 \qquad (5)$$

where $S_0$ represents the equilibrium signal obtained without applying the spin-lock pulse and $S(TSL_i)$ is the signal acquired at the *i*th TSL.

**2.4 Experimental setup**

The proposed neural network was implemented in TensorFlow with an NVIDIA GeForce RTX 2080 Ti GPU with the CUDA 10.1 library. For each dataset, the $T_{1\rho}$-weighted images from four volunteers were used for training, the images from another volunteer were used for validation and the images from the last three volunteers were used for testing. The 3D image volumes were split into 2D slices for training, validation and testing; e.g., there are total 2150 2D images (86 * 5 * 5) for training and validation, 1290 2D images (86 * 5 * 3) for testing.

First, the necessity of the generative module was evaluated. Here, the generative module was used as an independent network (G-Net). Its input was the TSL1 and TSL5 images, and the output was the generated synthetic TSL2, TSL3, and TSL4 images. The $T_{1\rho}$-weighted images from both the input and output were combined to estimate the $T_{1\rho}$ map. The fully sampled images were used for training and testing, and Loss$_2$ was used as the loss function for the generative network. The $T_{1\rho}$ map fitted with only two $T_{1\rho}$-weighted images (TSL1 and TSL5) was also calculated for comparison. In addition, the effect of the k-space noise level on the entire RG-Net and fitting process was also determined. Complex Gaussian white noise with different standard deviations was added into the input k-space data. The signal-to-noise ratios (SNRs) were 30dB, 25dB and 20dB. The SNR is calculated with the mean value of pixels in the region of interest (ROI) and the STD of the complex Gaussian white noise:

$$SNR = 20 * \log_{10} \frac{mean(m)}{std(g)} \qquad (6)$$

The noisy k-space data was trained and tested on RG-Net to observe the effect of noise on quality of $T_{1\rho}$-map.

The whole framework was then evaluated on the undersampled datasets. Training both the reconstruction module and generative module from scratch makes convergence difficult. Hence, the whole training process was divided into three sequential steps. In the first step, the reconstruction module was trained alone for 30 epochs, the corresponding k-space data of all training images were fed into the network, and the fully sampled images were used as the ground truth. $Loss_1$ was used for reconstruction module parameter updating. In the second step, the generative module was trained alone for 5 epochs, k-space data of TSL1 and TSL5 images were used as input, and the corresponding fully sampled TSL2, TSL3, TSL4 images were used as the ground truth. $Loss_2$ was used for generative module parameter updating. In the last step, the reconstruction module and generative module were trained together for 50 epochs, the k-space data of the TSL1 and TSL5 images were used together as input, the corresponding fully sampled TSL1 and TSL5 images were used as the ground truth for $Loss_1$, and the corresponding fully sampled TSL2, TSL3, and TSL4 images were used as the ground truth for $Loss_2$. $Loss_3 = Loss_1 + \lambda * Loss_2$ was used for RG-Net parameter updating. We set $\lambda = 0.1$ empirically due to the best performance in our work. The network was optimized using the ADAM optimizer with a fix learning rate of 0.0005. Since only two undersampled k-space were used as input in our method, the acceleration factor in the parametric direction ($R_{TSL}$) was 2.5 relative to the total 5 acquired $T_{1\rho}$-weighted images in usual. The net acceleration rate ($R_e$) was $R_K \times R_{TSL}$. For comparison, the ADMM-Net III and deep MANTIS network [27] were used to reconstruct five $T_{1\rho}$ weighted images from highly undersampled k-space ($R_{TSL} = 1$) or two $T_{1\rho}$ weighted images from moderately undersampled k-space ($R_{TSL} = 2.5$) with the same $R_e$ as that in RG-Net. To jointly reconstruct all 5 contrasts when RTSL = 1, undersampled k-space data of images with all 5 TSLs were input into ADMM-Net III in their entirety. 3D Convolution was also used in ADMM-Net III when $R_{TSL} = 1$.. The deep MANTIS framework has recently been applied for fast MR parametric mapping to improve the estimation accuracy. As a replacement for traditional pipeline of reconstruction followed by pixel-wise curve fitting, it directly maps zero-filled images to parametric map under the supervision of fully sampled datasets.. In our experiment, the MANTIS was trained and tested on the same data as R-Net and RG-Net. The zero-filled images were feed into the U-shaped network. The loss of the MANTIS network consists of two parts. The first part ensures that the reconstructed parameter maps from the network mapping produce synthetic undersampled k-space matching the acquired k-space measurements in the k-space domain. The second part ensures that the undersampled images produce parameter maps that are the same as the reference parameter maps generated from fully sampled images. The network was optimized using the ADAM optimizer with a fixed learning rate of 0.0002 for a total of 400 epochs. Li et al[33] proposed a deep learning method to directly estimate the $T_{1\rho}$ maps from undersampled k-space of reduced echoes. This method was also used as a comparison. The network was optimized using the ADAM optimizer with a fixed learning rate of 0.0002 for 400 epochs.

To evaluate the performance of G-Net on standard compressed sensing reconstruction, we chose the L+S method for reconstruction. We used L+S to reconstruct TSL1 and TSL5 images at $R_k = 6.8$ and G-Net to generate TSL2,3 and 4 images. All five TSL images were reconstructed at $R_k = 17$ for comparison.

To evaluate the ability of reliably identifying pathology, a dataset from a patient with glioma is used for testing. This dataset is acquired at $R_K = 3$ and reconstructed on UIH 3.0T scanner uCS system in consideration of the patient's physical condition. The G-Net was trained with reconstructed images from volunteers at $R_K = 3$ and then evaluated with images from the patient.

The quality of the generated synthetic images and the estimated $T_{1\rho}$ maps were evaluated by nRMSE within manually drawn ROIs [36]:

$$\text{nRMSE} = \sqrt{\frac{\|x_{est} - x_{ref}\|_2^2}{\|x_{ref}\|_2^2}} \quad (7)$$

where $x_{est}$ denotes the reconstructed/generated image or $T_{1\rho}$ map fitted from these images and $x_{ref}$ is the fully sampled image or $T_{1\rho}$ map fitted from them.

## 3 | RESULTS

Figure 2 shows the $T_{1\rho}$ maps estimated from 2 acquired images plus 3 generated images from the generative module. Compared with the $T_{1\rho}$ maps estimated from only two acquired images using pixel-wise fitting, the $T_{1\rho}$ maps of our method have a lower nRMSE value and smaller errors according to the error maps.

Figure 3 shows the $T_{1\rho}$ maps of the brain at $R_e = 11.5$ and 17 for the 6 methods including RG-Net, R-Net (ADMM-Net III, $R_{TSL} = 1$ and 2.5), MANTIS ($R_{TSL} = 1$ and 2.5) and MSCNN ($R_{TSL} = 2.5$). The $T_{1\rho}$ map of RG-Net had the lowest nRMSE value among the five strategies. At the same net acceleration factor, the nRMSE values of ADMM-Net III and MANTIS were higher than those of the other two methods that involve undersampling in the parametric direction. At a higher acceleration factor of $R_e = 17$, the $T_{1\rho}$ map from ADMM-Net III and MANTIS with $R_{TSL} = 1$ showed obvious aliasing or blurring artifacts due to the high undersampling of the k-space.

Figure 4 shows the synthetic $T_{1\rho}$-weighted image at TSL = 10 ms from the generative module compared with the image reconstructed by ADMM-Net III from the knee dataset with $R_k = 17$. The $T_{1\rho}$-weighted image reconstructed by ADMM-Net III is very blurred due to the high undersampling of the k-space data. The nRMSE value of ADMM-Net III is therefore higher than that of the synthetic images.

Figure 5 shows the $T_{1\rho}$ maps overlaid on the corresponding $T_{1\rho}$-weighted image at $R_e = 17$ from five methods. RG-Net has the lowest nRMSE value of the $T_{1\rho}$ map, and the quality of the $T_{1\rho}$ map has been improved with the help of a generative module. The findings are similar to those in Figure 3.

Figure 6(a) shows the synthetic $T_{1\rho}$-weighted image at TSL = 10 ms from the generative module based on the L+S reconstructed $T_{1\rho}$-weighted image compared with the image reconstructed by L+S from the knee dataset with $R_k = 17$. The nRMSE value of the synthetic image from G-Net & L+S is lower than that of L+S reconstruction alone. Figure 6(b) shows the $T_{1\rho}$ maps overlaid on the corresponding $T_{1\rho}$-weighted image at $R_e = 17$ from four methods. Except for RG-Net, the G-Net & L+S hads the lowest nRMSE value of the T1ρ map. The strategy of both undersampling in the k-space and reducing the contrast number also shows improvement on improved the T1ρ map.

Figure 7 shows brain $T_{1\rho}$ maps of a patient with glioma. The $T_{1\rho}$ map from five uCS reconstructed $T_{1\rho}$-weighted images is regarded as the ground Truth. Due to the contribution of generative module, the $T_{1\rho}$ map from two uCS reconstructed $T_{1\rho}$-weighted images and three synthetic images was more accurate than that from two uCS reconstructed images alone.

Figure S1 shows the comparison between the T1ρ maps from RG-Net and R-Net under different SNRs at $R_k$ = 6.8. According to the nRMSE curves obtained by RG-Net and R-Net at different SNRs in Figure S2, the nRMSE value of the T1ρ map decreases as the SNR increases, but the overall nRMSE of the RG-Net is smaller than that obtained by R-Net.

Figure S3 shows the statistical results of $T_{1\rho}$ maps on 3 testing subjects from 6 methods in brain and knee datasets. According to the boxplot, our method achieved lowest nRMSE among all the comparison methods. The strategy of both undersampling in k-space and reducing contrast number also show improvement on quality of $T_{1\rho}$ maps.

Figure S4 shows the training loss and validation loss in the last step of training process. The validation loss didn't increase during the training process, which indicates that our method is not over-fitting to the training data.

## 4 | DISCUSSION and CONCLUSION

In this work, we developed a fast MR parametric mapping method by undersampling in the parametric direction as well as in the k-space simultaneously. The generative module makes use of the process of $T_{1\rho}$ signal relaxation. The undersampling operation in the parametric direction reduces the need for a high acceleration factor in the k-space when the net acceleration rate is fixed. With the help of the generative module, the adverse effect of undersampling in the parametric direction is weakened since the lost information in the parametric direction is mostly recovered. All this leads to lower reconstruction errors on contrast images and a more accurate $T_{1\rho}$ estimation map than those obtained by methods only undersampling in the k-space.

In the proposed framework, the exponential model is implicitly learned by the generative module through the supervision of the fully sampled multi-contrast images. The generative module can learn the prior correlation among images from the training data.

The RG-Net also shows superior performance in terms of the $T_{1\rho}$ map compared to R-Net with different noise levels at $R_k = 6.8$. The quality of the corresponding $T_{1\rho}$ map decreases with a decreasing SNR, but the decline in quality of RG-Net is smaller than that of R-Net (see Supporting Information Figure S1 and Figure S2).

The generative network in our framework can be easily integrated into other fast parametric mapping methods to achieve a higher acceleration factor. Taking the low rank plus sparse [22] (L+S) method as an example, we used the L+S method to reconstruct the TSL1 and TSL5 images from the undersampled k-space data followed by the generative network to generate TSL2, TSL3, and TSL4 images. The results are shown in Figure 6. We can see that when the generative module is used, it improves the $T_{1\rho}$ map quality compared with only the reconstructed images of L+S at a high acceleration factor being used.

Our method's ability of reliably identify pathology was also evaluated. The G-Net was trained with images from volunteers and evaluated with images from a patient with glioma. The error of $T_{1\rho}$ map with G-Net is slightly lower than that of $T_{1\rho}$ map without G-Net, which shows that our method can generalize to abnormal cases. Due to the limitation of acquisition coils, the images were reconstructed using the CS method provided by the vendor rather than using the trained RG-Net. Due to the limitation of acquisition coils, the images were reconstructed using the CS method provided by the vendor rather than using the trained RG-Net. Due to this limitation, the generalizability of RG-Net to abnormal cases is unclear. However, the generalizability of G-Net to abnormal cases was evaluated through the experiment. Our method improved the quality of the T1ρ map for the patient with glioma without being trained on abnormal for two reasons. First, the mapping of G-Net is learned by the MR signal decay model of healthy subjects, which is consistent with the signal model in images from the patient. Second, the T1ρ value from some parts of healthy subjects, such as cerebrospinal fluid, is close to that from the lesion site of abnormal cases.

In this study, the $T_{1\rho}$ map was estimated with the traditional pixel-wise least squares curve fitting method. In previous studies, a deep neural network named MANTIS was applied to direct estimate parametric map from contrast images and shown an improved estimation accuracy[27]. In our experiment, MANTIS yields better performance than R-Net with pixel-wise curve fitting at $R_k = 17$, which shows the strong potential of direct mapping methods in MR parametric mapping using deep learning. Although the strategy of decreasing the contrast number shows improved performance in MANTIS, this improvement is less obvious than those of using ADMM-Net III and L+S combined with the generative module (see Figures 5 and 6), which can recover the information loss in the contrast dimension.

This study has several limitations. The training of the generative module requires a large number of fully sampled multi-contrast images to fully incorporate the signal model into the image generation, where the acquisition time of the 3D volume parametric mapping data is very long[7]. Due to GPU limitations, the generative module is an end-to-end convolutional neural network trained with 2D rather than 3D slices of $T_{1\rho}$ contrast images, so it may be sensitive to through-plane motion artifacts.

In conclusion, we propose a novel fast MR parametric mapping method based on deep learning by undersampling both in the parametric direction and the k-space data simultaneously. Our method can generate synthetic multi-contrast images from acquired images under the constraints of the $T_{1\rho}$ signal relaxation model and the correlation between different contrast images. Our method yields better performance than the methods that undersample the k-space data only at the same net acceleration factor. The generative module can also be used as an insert module in other fast MR parametric mapping methods to achieve a higher acceleration factor.

**ACKNOWLEDGEMENTS**

Funding: This work was partially supported by the National Key R&D Program of China Nos. 2017YFC0108800, 2020YFA0712200, National Natural Science Foundation of China under grant Nos. 61771463, 81830056, U1805261, 61671441

and 81971611, the Innovation and Technology Commission of the government of Hong Kong SAR under grant no. MRP/001/18X, the Key Laboratory for Magnetic Resonance and Multimodality Imaging of Guangdong Province under grant No. 2020B1212060051, and by the Chinese Academy of Sciences program under grant No. 2020GZL006.

**LIST OF TABLES**

Table 1: The learnable parameters and time usage of different networks.

| Method | Params (K) | Training Time (s/epoch) | Inference Time (s/epoch) |
|---|---|---|---|
| R-Net | 161 | 358.16 | 134.72 |
| RG-Net | 237 | 397.42 | 162.53 |
| MANTIS | 120 | 17.49 | 7.54 |
| MSCNN | 638 | 23.65 | 9.86 |

**FIGURE CAPTIONS**

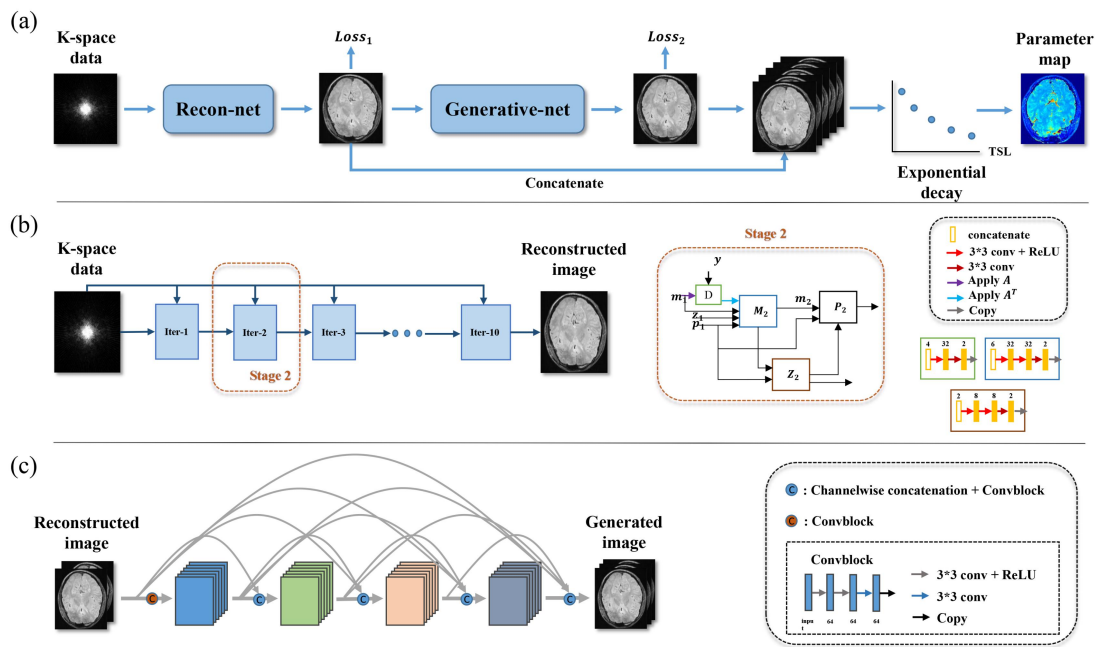

Fig. 1: (a) The proposed MR parametric mapping method consists of two modules. (b) The reconstruction module uses ADMM-Net III to reconstruct images from undersampled k-space data. (c) The generative module is a densely connected neural

network that generates the desired contrast-weighted images from few reconstructed images under the supervision of the corresponding weighted images.

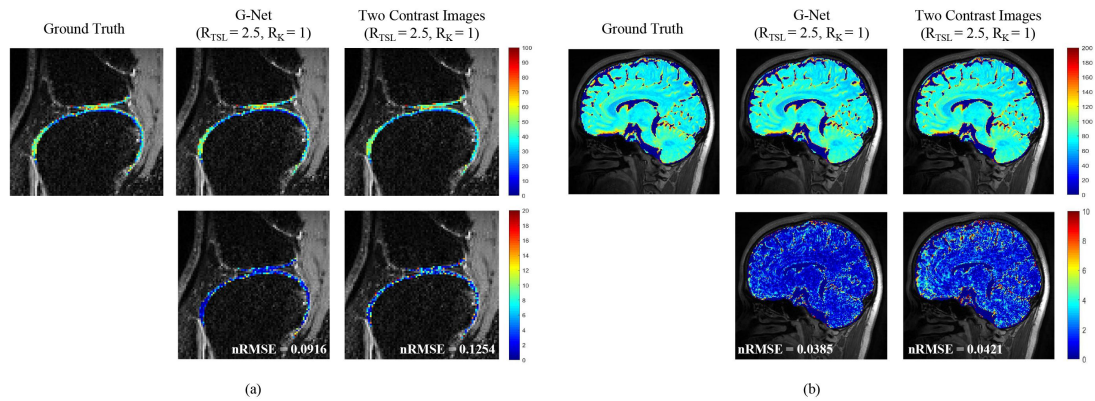

Fig. 2: Estimated $T_{1\rho}$ maps and error maps obtained by G-Net and two contrast images without acceleration. The $T_{1\rho}$ maps obtained by G-Net exhibit better agreement with the ground truth. (a) Analysis of $T_{1\rho}$ maps from knee. (b)Analysis of $T_{1\rho}$ maps from brain.

(b) nRMSE values of the brain $T_{1\rho}$ maps at different noise levels (the STD of noise). The decline in the quality of the $T_{1\rho}$ maps obtained by G-Net is smaller than that obtained by the two contrast images according to the nRMSE analysis. G-Net: The $T_{1\rho}$ map from two fully sampled images and three synthetic images. Two contrast images: The $T_{1\rho}$ map from only two contrast images.

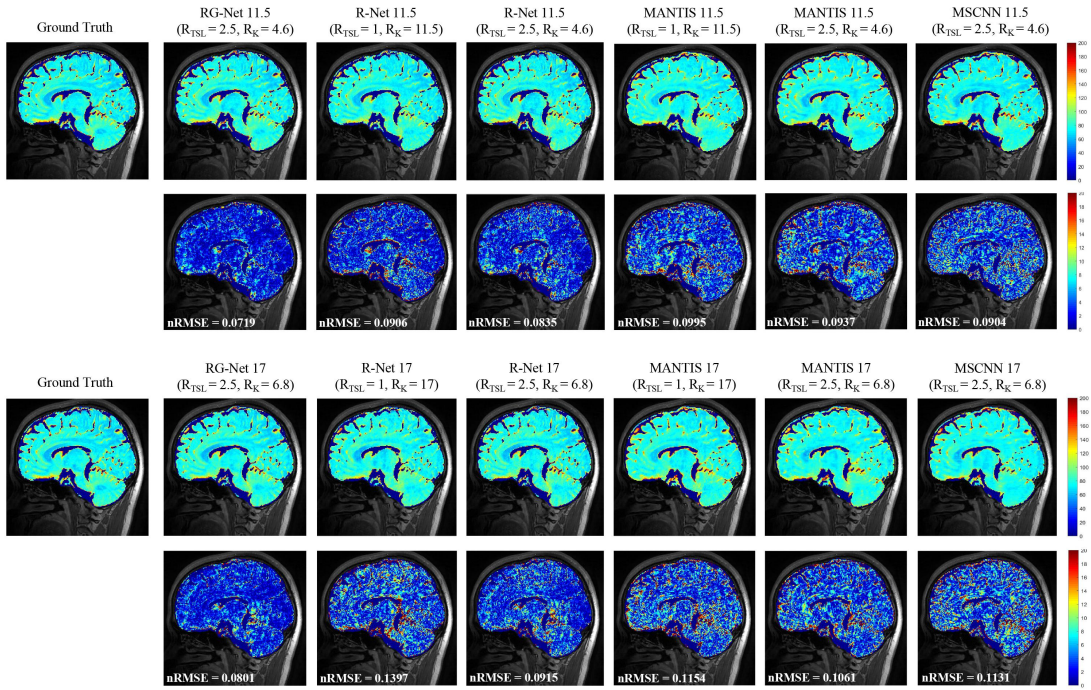

Fig. 3: Estimated $T_{1\rho}$ maps and error maps obtained by RG-Net, R-Net with different strategies, MANTIS and MSCNN with net acceleration factors of $R_e = 11.5$ and 17 from retrospectively undersampled data. The $T_{1\rho}$ maps obtained from RG-Net exhibit the best agreement with the ground truth. The strategy of undersampling in two dimensions is shown to be effective by the results of R-Net. RG-Net: Mapping with 2 acquired weighted images and 3 synthetic images. R-Net: ($R_{TSL} = 1$) Mapping with five acquired images. R-Net ($R_{TSL} = 2.5$): Mapping with two acquired images. MANTIS ($R_{TSL} = 1$): The results of the deep MANTIS network with five acquired images. MANTIS ($R_{TSL} = 2.5$): The results of the deep MANTIS network with two acquired images. MSCNN : The results of MSCNN with two acquired images.

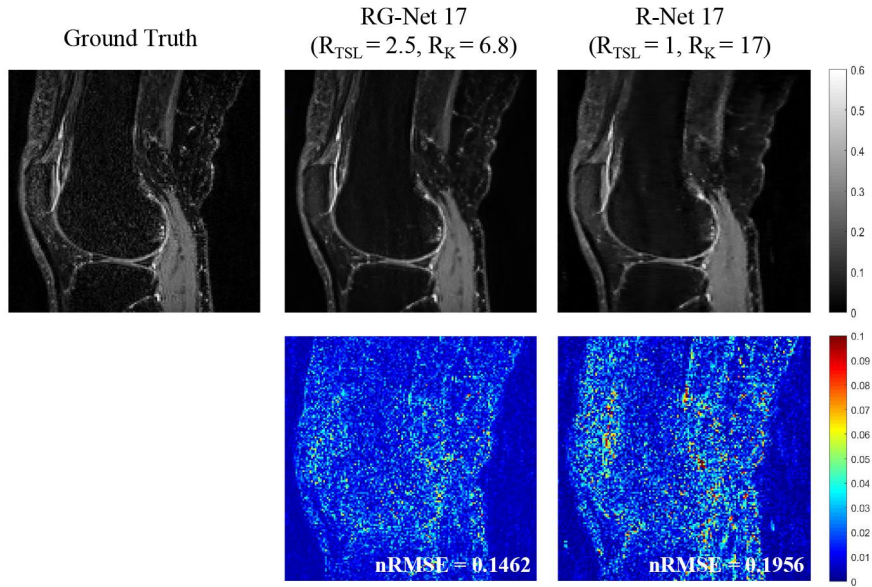

Fig. 4: Weighted images and error maps from the generative module and reconstruction module with the same net acceleration rate $R_e = 17$. The synthetic weighted image has a lower nRMSE than the weighted images from the reconstruction module. RG-Net 17: The synthetic images generated from the results of the reconstruction module at $R_K = 6.8$. R-Net 17: The result of the reconstruction module at $R_K = 17$.

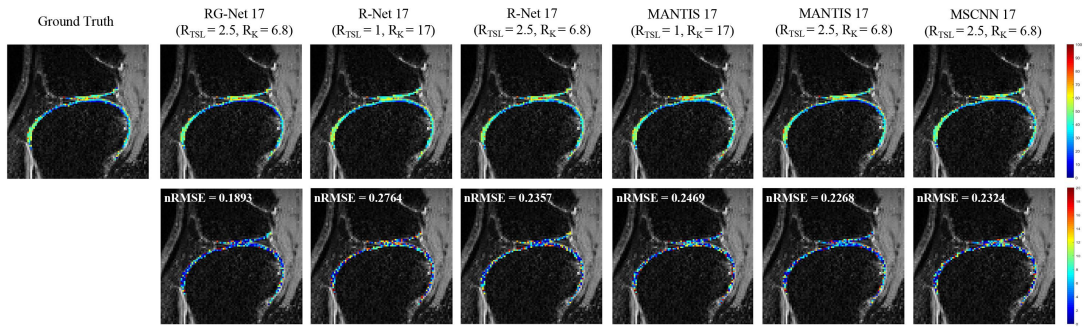

Fig. 5: Estimated $T_{1\rho}$ maps and error maps obtained by RG-Net, R-Net with different strategies, MANTIS and MSCNN with a net acceleration factor of $R_e = 17$ from retrospectively undersampled data. The $T_{1\rho}$ maps obtained from RG-Net exhibit the best agreement with the ground truth. The effectiveness of the undersampling strategy in two dimensions is shown by the results of R-Net. RG-Net: Mapping with 2 acquired weighted images and 3 synthetic images. R-Net: ($R_{TSL} = 1$) Mapping with five acquired images. R-Net ($R_{TSL}$ = 2.5): Mapping with two acquired images. MANTIS ($R_{TSL}$ = 1): The results of the deep MANTIS network with five acquired images. MANTIS ($R_{TSL}$ = 2.5): The results of the deep MANTIS network with two acquired images. MSCNN : The results of MSCNN with two acquired images.

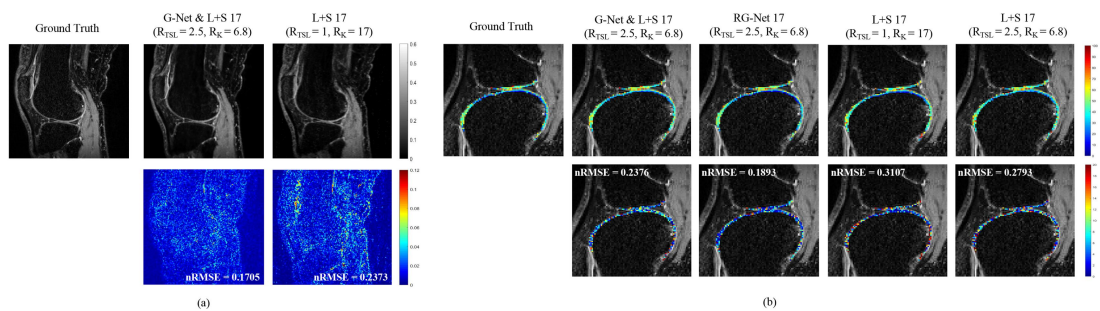

Fig. 6: The results of the proposed method based on L+S reconstruction;

(a) Synthetic weighted images from the generative module based on L+S reconstruction and weighted images from conventional L+S with the same net acceleration rate $R_e$ = 17. The synthetic weighted image has a lower nRMSE than the reconstructed weighted images from L+S. G-Net & L+S 17: The synthetic images generated from the results of L+S at $R_K$ = 6.8 and $R_{TSL}$ = 2.5. L+S 17: The result of L+S reconstruction at $R_K$ = 17 and $R_{TSL}$ = 1.

(b) Estimated $T_{1\rho}$ maps and error maps obtained by G-Net & L+S, RG-Net and L+S with different strategies with a net acceleration factor of $R_e$ = 17 from retrospectively undersampled data. Except for RG-Net, the $T_{1\rho}$ maps obtained from G-Net & L+S exhibit the best agreement with the ground truth. The effectiveness of the strategy of undersampling in two dimensions is shown by the results of L+S. G-Net & L+S 17: Mapping with 2 acquired weighted images from L+S reconstruction at $R_K$ = 6.8 and $R_{TSL}$ = 2.5 and with 3 synthetic images. RG-Net 17: Mapping with 2 acquired weighted images at $R_K$ = 6.8 and 3 synthetic images. L+S 17 ($R_{TSL}$ = 1, $R_k$=17): Mapping with 5 acquired weighted images from L+S at $R_K$ = 17. L+S 17X ($R_{TSL}$ = 2.5, $R_k$=6.8): Mapping with 2 acquired weighted images from L+S at $R_K$ = 6.8.

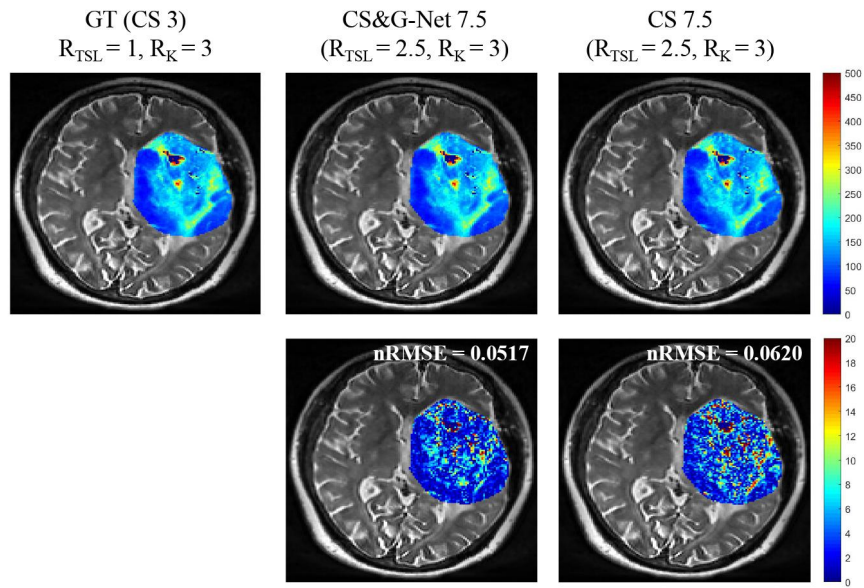

Fig. 7: Estimated $T_{1\rho}$ maps and error maps obtained by CS&G-Net and CS with a net acceleration factor of $R_e$ = 7.5 from prospective undersampled data. The $T_{1\rho}$ maps obtained from CS&G-Net exhibit better agreement with the ground truth than CS reconstruction alone. GT(CS3): Mapping with five weighted images reconstructed using UIH uCS at $R_k$ = 3. CS&G-Net 7.5: Mapping with two reconstructed weighted images using UIH uCS at $R_k$ = 3 and three synthetic images. CS 7.5: Mapping with two reconstructed images using UIH uCS at $R_k$ = 3.

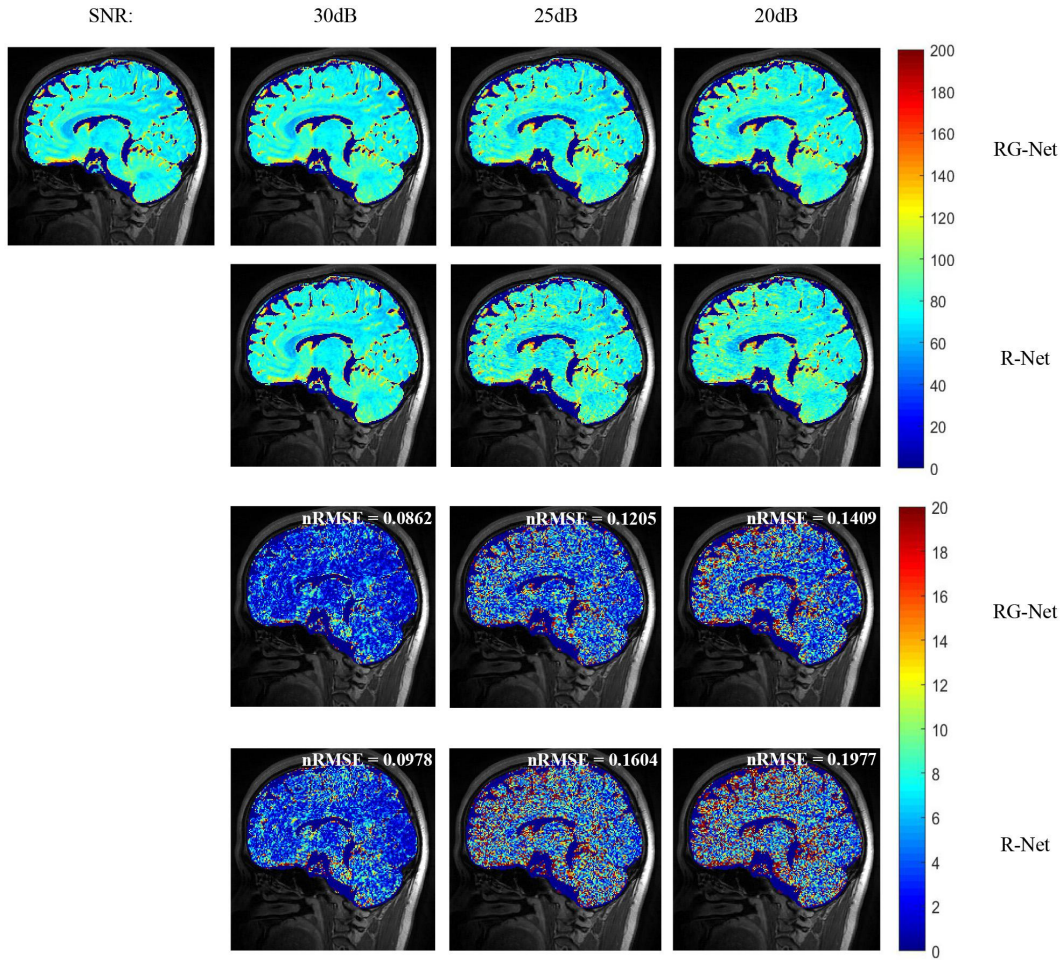

Supporting Information Fig. S1: Estimated $T_{1\rho}$ maps and error maps obtained by RG-Net and two contrast images at $R_k = 6.8$ with different SNRs. RG-Net: The $T_{1\rho}$ map from two reconstructed images at $R_k = 6.8$ and three synthetic images. Two contrast images: The $T_{1\rho}$ map from only two reconstructed images at $R_k = 6.8$.

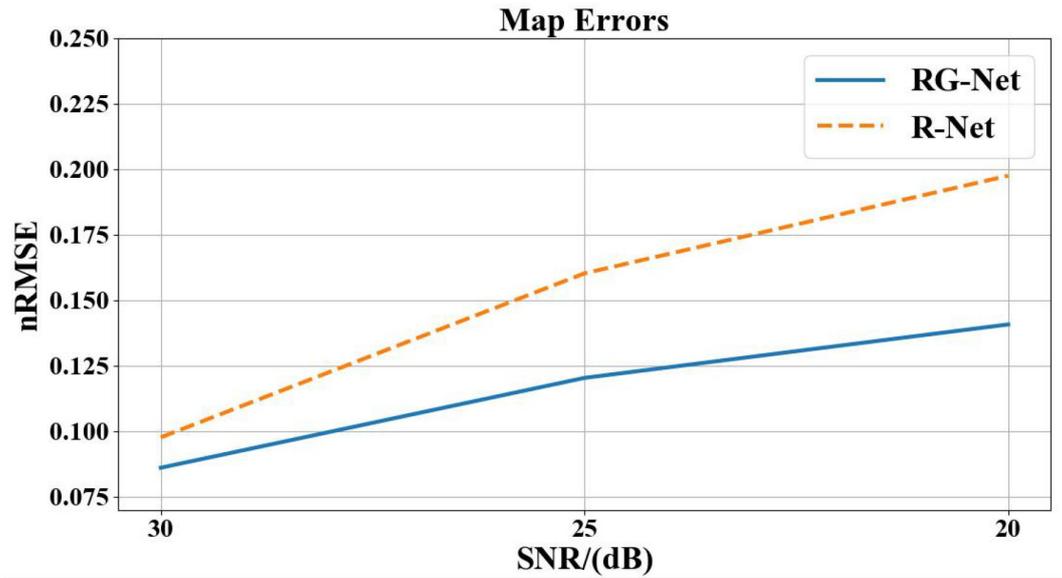

Supporting Information Fig. S2: The correlation between the SNR and nRMSE values at $R_k = 6.8$.

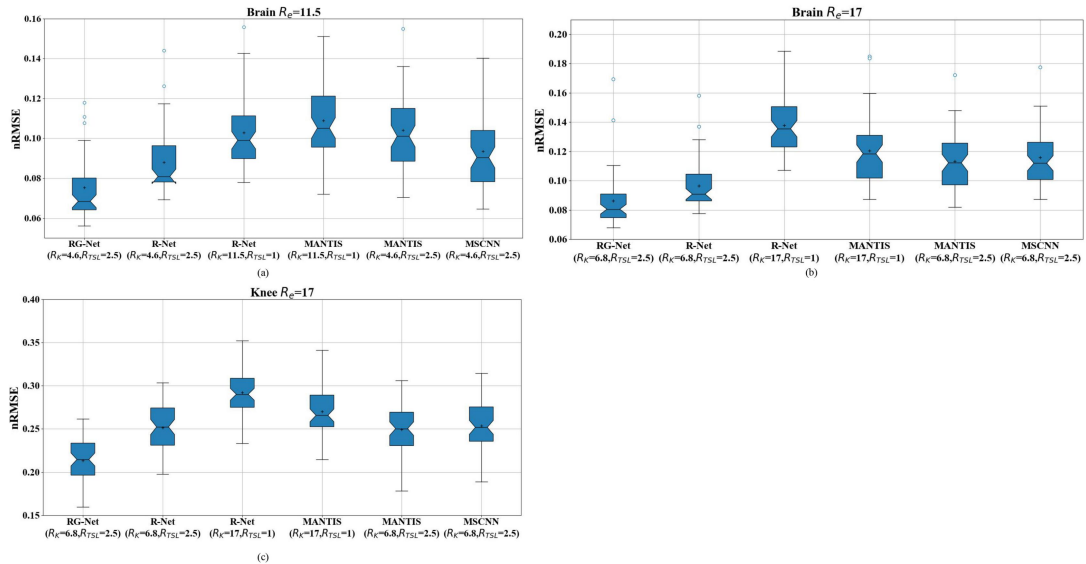

Supporting Information Fig. S3: The statistical results including the mean, median and quartile of $T_{1\rho}$ maps from 6 methods on brain and knee datasets at the same acceleration factor. (a) $R_e$=11.5 for brain. (b) $R_e$ = 17 for brain. (c) $R_e$ = 17 for knee.

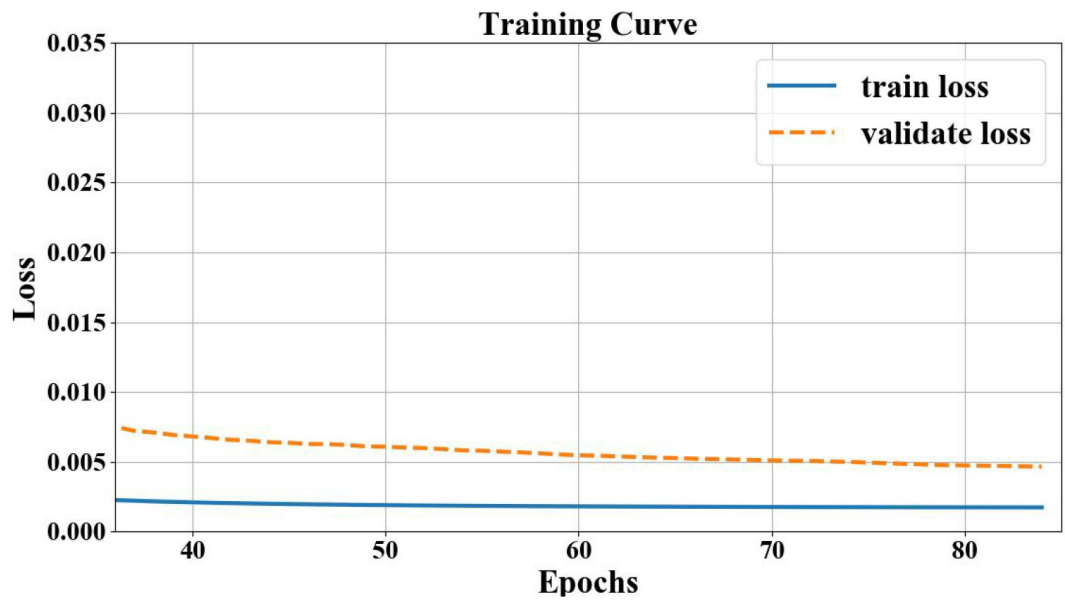

Supporting Information Fig. S4: The training loss and validation loss curve in the last step of the training process.